\documentclass[10pt,twocolumn,aps,prl,nofootinbib,preprintnumbers,groupedaddress,floatfix]{revtex4} %ADF13: added floatfix
\usepackage[latin9]{inputenc}
\usepackage{bm}
\usepackage{amsmath}
\usepackage{amssymb}
\usepackage{graphicx}

\makeatletter
\@ifundefined{textcolor}{}
{%
 \definecolor{BLACK}{gray}{0}
 \definecolor{WHITE}{gray}{1}
 \definecolor{RED}{rgb}{1,0,0}
 \definecolor{GREEN}{rgb}{0,1,0}
 \definecolor{BLUE}{rgb}{0,0,1}
 \definecolor{CYAN}{cmyk}{1,0,0,0}
 \definecolor{MAGENTA}{cmyk}{0,1,0,0}
 \definecolor{YELLOW}{cmyk}{0,0,1,0}
}

\usepackage{bm}%\usepackage{esint}

\def\Mpl{M_{\rm G}}
\def\MPl{\tilde M_{\rm G}}

\newcommand{\lmk}{\left(}
\newcommand{\rmk}{\right)}

\newcommand{\cM}{{\cal M}}
\newcommand{\ts}{{\tilde{s}}}
\newcommand{\tg}{{\tilde{g}}}

\newcommand{\tc}{{\tilde{c}}}
\newcommand{\ta}{{\tilde{a}}}
\newcommand{\tR}{{\tilde{R}}}
\newcommand{\tB}{{\tilde{B}}}

\newcommand{\lamg}{\lambda_\mu}

\makeatother

\begin{document}

\title{Possible existence of viable models of bi-gravity with detectable
graviton oscillations by gravitational wave detectors}

\author{Antonio De Felice}

\affiliation{ThEP's CRL, NEP, The Institute for Fundamental Study, Naresuan University,
Phitsanulok 65000, Thailand}

\affiliation{Thailand Center of Excellence in Physics, Ministry of Education,
Bangkok 10400, Thailand}

\author{ Takashi Nakamura}

\affiliation{Department of Physics, Kyoto University Kyoto 606-8502, Japan }

\author{Takahiro Tanaka}

\affiliation{Yukawa Institute for Theoretical Physics, Kyoto, Japan}
\begin{abstract}
We discuss graviton oscillations %ADF: plural
 based on the ghost free bi-gravity theory. We point out that this
theory possesses a natural cosmological background solution which
is very close to the case of general relativity. Furthermore, interesting
parameter range of the graviton mass, which can be explored by the
observations of gravitational waves, is not at all excluded by the
constraint from the solar system tests. Therefore the graviton oscillation
with possible inverse chirp signal would be an interesting scientific
 target of KAGRA, adv LIGO, adv Virgo and GEO. 
\end{abstract}

\date{\today}

\maketitle
\textbf{Introduction:} %\textbf{The model and the background:}
Many works have been done for the detection possibility of modified
propagation of gravitational waves due to finite graviton mass~\cite{Will:1997bb,Yagi:2009zm}.
However, adding mass to graviton was thought to be theoretically problematic
due to the so-called Boulware-Deser (BD) ghost~\cite{Boulware:1973my}.

Recently, Hassan and Rosen proposed the first example of ghost-free
%ADF: removed "a"
bi-gravity models~\cite{Hassan:2011zd}, based on the fully nonlinear
massive gravity theory in which the %ADF13:added "the"
 Boulware-Deser ghost is removed by
construction~\cite{deRham:2010ik,deRham:2010kj,Hassan:2011hr}. We
consider two metrics %ADF: changed plural
expressed by 
\begin{eqnarray*}
ds^{2} & = & g_{\mu\nu}dx^{\mu}dx^{\nu}\,,\quad d\ts^{2}=\tg_{\mu\nu}dx^{\mu}dx^{\nu}\,.
\end{eqnarray*}
We introduce a ghost free action $S=\int d^{4}x\mathcal{L}$ with
\begin{eqnarray*}
\mathcal{L} & = & \sqrt{-g}\left[\Mpl^{2}\left(\frac{R}{2}-m^{2}\sum_{n=0}^{4}c_{n}V_{n}(Y_{\nu}^{\mu}),\right)+L_{{\rm m}}\right]\\
 &  & \qquad+{\frac{\kappa\Mpl^{2}}{2}}\sqrt{-\tg}\tR\,,
\end{eqnarray*}
where $\Mpl^{2}=1/(8\pi G_{N})$; $G_{N}$ is the gravitational constant;
$Y_{\nu}^{\mu}=\sqrt{g^{\mu\alpha}\tg_{\alpha\nu}}$; $R$ and $\tilde{R}$
are the Ricci scalars with respect to $g_{\mu\nu}$ and $\tg_{\mu\nu}$,
respectively; $g$ and $\tg$ are the determinants of $g_{\mu\nu}$
and $\tg_{\mu\nu}$, respectively; $\kappa$ is a constant which expresses
the ratio between the two gravitational constants for $\tg_{\mu\nu}$
and $g_{\mu\nu}$; $c_{n}$ ($n=0,\dots,4$) are dimensionless constants,
and $L_{{\rm m}}$ is the Lagrangian of the matter which interacts
only with $g_{\mu\nu}$. %ADF: a few changes in this paragraph
By expressing the trace of $Y^{n}$ as $[Y^{n}]={\rm tr}(Y^{n})=Y_{\alpha_{1}}^{\alpha_{0}}Y_{\alpha_{2}}^{\alpha_{1}}\cdots Y_{\alpha_{0}}^{\alpha_{n-1}},$
we can write $V_{n}$'s as 
\begin{eqnarray*}
V_{0} & = & 1~,\quad V_{1}=[Y],\quad V_{2}=[Y]^{2}-[Y^{2}]~,\\
V_{3} & = & [Y]^{3}-3[Y][Y^{2}]+2[Y^{3}]~,\\
V_{4} & = & [Y]^{4}-6[Y]^{2}[Y^{2}]+8[Y][Y^{3}]+3[Y^{2}]^{2}-6[Y^{4}]~.
\end{eqnarray*}
The variation of the action with respect to $g^{\mu\nu}$ and $\tg^{\mu\nu}$
yields the field equations as 
\begin{eqnarray*}
R_{\mu\nu}-\frac{1}{2}g_{\mu\nu}R+B_{\mu\nu} & = & \Mpl^{-2}T_{\mu\nu},\\
\kappa\left[\tR_{\mu\nu}-\frac{1}{2}\tg_{\mu\nu}\tR\right]+\tB_{\mu\nu} & = & 0,
\end{eqnarray*}
where $T_{\mu\nu}$ is the energy momentum tensor of the ordinary
matter, whereas %ADF: and -> whereas
$B_{\mu\nu}$ and $\tB_{\mu\nu}$ come from the variations of the
mass term. $B_{\mu\nu}$ and $\tB_{\mu\nu}$, as well as $T_{\mu\nu}$,
satisfy conservation laws, which are explicitly given by 
\begin{equation}
\nabla_{\mu}B_{\nu}^{\mu}=0\,,\quad\nabla_{\mu}T_{\nu}^{\mu}=0\,,\quad\tilde{\nabla}_{\mu}\tB_{\nu}^{\mu}=0\,,\label{cons}
\end{equation}
%which are consequences of the Bianchi identity, from the divergence
%of Eq.\ (10) and (11), and 
where $\nabla$ and $\tilde{\nabla}$ are the covariant derivative
operators %ADF: changed here
 with respect to $g_{\mu\nu}$ and $\tg_{\mu\nu}$, respectively.

\textbf{The cosmological background:~} The background cosmology of
this theory has been widely studied in Refs.~\cite{Volkov:2011an,Comelli:2011zm,Comelli:2012db},
but here our focus is on a particularly healthy branch. We assume
that the two metrics %ADF: metrics
 can be written as 
\begin{eqnarray*}
ds^{2}=a^{2}(-dt^{2}+d\bm{x}^{2})\,,\quad d\tilde{s}^{2}=\ta^{2}(-\tc^{2}dt^{2}+d\bm{x}^{2})\,,
\end{eqnarray*}
where $a$, $\ta$, and $\tc$ are functions of the time coordinate
$t$. The Friedmann equation for the physical metric reads %\begin{widetext}
\begin{eqnarray}
3H^{2}=\frac{\rho_{{\rm m}}+\rho_{V}}{\Mpl^{2}}\,,\label{eq:fried1}
\end{eqnarray}
where we have introduced the Hubble parameter $H\equiv\dot{a}/a^{2}$,
the matter energy density $\rho_{{\rm m}}$ including the dark energy,
and the energy density due to the mass term 
\begin{eqnarray*}
\rho_{V}(\xi)\equiv\Mpl^{2}m^{2}(c_{0}+3\xi c_{1}+6\xi^{2}c_{2}+6\xi^{3}c_{3})\,,
\end{eqnarray*}
with $\xi\equiv\ta/a$. The Friedmann equation for the hidden metric
reads as follows %ADF: small changes
\begin{equation}
\frac{3}{\tc^{2}a^{2}}\left(\frac{\dot{\ta}}{\ta}\right)^{2}=\frac{m^{2}}{\kappa}\left(\frac{c_{1}}{\xi}+6c_{2}+18\xi c_{3}+24\xi^{2}c_{4}\right),\label{eq:fried2}
\end{equation}
%\end{widetext}

%Multiplying by $a$ both sides of the first Friedmann equation, Eq.\ (\ref{eq:fried1}),
%and taking the first time derivative, we have
Writing down the first equation in Eq.~(\ref{cons}), we have 
\[
3\,\Gamma(\xi)\,[\tc aH-(\dot{\ta}/{\ta})]=0\,,
\]
where $\Gamma(\xi)\equiv c_{1}\xi+4c_{2}\xi^{2}+6c_{3}\xi^{3}$. This
equation can be solved by imposing $\Gamma(\xi)=0$ or $\tc aH-(\dot{\ta}/{\ta})=0$,
%(excluding the possibility of $\xi=0$)
which implies the existence of two branches. In the following we will
discuss the \emph{physical branch}, defined by the latter condition,
since the other branch is pathological%
\footnote{The degrees of freedom of the theory reduce~\cite{Comelli:2012db}.
This will lead to a similar phenomenology observed in the original
ghost-free single-metric massive gravity, which is characterized by
the presence of a scalar non-perturbative ghost~\cite{DeFelice:2012mx}.%ADF: introduced citation
}. Combining this condition with Eqs.~(\ref{eq:fried1}) and (\ref{eq:fried2}),
we obtain an algebraic equation for $\xi$ 
\begin{eqnarray}
\frac{\rho_{{\rm m}}}{\Mpl^{2}m^{2}} & = & \Biggl[\frac{c_{1}}{\kappa\xi}+\left(\frac{6c_{2}}{\kappa}-c_{0}\right)+\left(\frac{18c_{3}}{\kappa}-3c_{1}\right)\xi\nonumber \\
 &  & \qquad+\left(\frac{24c_{4}}{\kappa}-6c_{2}\right)\xi^{2}-6c_{3}\xi^{3}\Biggr]\,.\label{eq:algebraic}
\end{eqnarray}
If $m^{2}\gg\rho_{{\rm m}}/\Mpl^{2}$, the r.h.s of Eq. (4) should
be very small. %If we assume that the fundamental mass scale of the bi-gravity theory
%is sufficiently high, each term on the right hand side should be much
%larger than the left hand side $\rho_{{\rm m}}/\Mpl^{2}$ in the directly
%observable universe at a low energy regime. 
Denoting a value of $\xi$ at which the right hand side vanishes by
$\xi_{c}$, we focus on a cosmological background solution for which
$\xi$ asymptotes to $\xi_{c}$ for $\rho_{{\rm m}}\to0$. As we can
absorb the constant part of $\rho_{V}(\xi)$ into the cosmological
constant in $\rho_{{\rm m}}$, we also %ADF: also
assume that $\rho_{V}(\xi_{c})=0$.

For this type of solution, we can expand $\xi$ around $\xi_{c}$
at low energies. Keeping only the linear order in $\xi-\xi_{c}$,
Eq.~(\ref{eq:algebraic}) becomes %ADF: Eq.
\[
\frac{\xi-\xi_{c}}{\xi_{c}}\approx-\frac{\rho_{{\rm m}}}{3m^{2}\Mpl^{2}\Gamma_{c}}\frac{\kappa\xi_{c}^{2}}{1+\kappa\xi_{c}^{2}}\,,
\]
where $\Gamma_{c}\equiv\Gamma(\xi_{c})$. Substituting this relation
into Eq.~(\ref{eq:fried1}), we recover the usual Friedmann equation
as 
\begin{eqnarray*}
3H^{2}\approx{\MPl^{-2}}{\rho_{{\rm m}}}\,,
\end{eqnarray*}
with the effective gravitational constant given by 
\[
\MPl^{2}\equiv\Mpl^{2}{(1+\kappa\xi_{c}^{2})}\,.
\]

On using the definition of $\xi$, %ADF: On using
 the relation $\tc aH=\dot{\ta}/\ta$ implies $\dot{\xi}=(\tc-1)aH\xi$.
Substituting the differentiation of Eq.~(\ref{eq:algebraic}) into
this relation, we obtain 
\[
\tc\approx1+\frac{\kappa\xi_{c}^{2}(\rho_{{\rm m}}+P_{{\rm m}})}{\Gamma_{c}m^{2}\MPl^{2}}\,,
\]
at low energies, where $P_{{\rm m}}$ is the matter pressure density.
The above relation implies that the light cone of the hidden metric
automatically gets closer to the physical one as the matter energy
density is diluted.

\textbf{Propagation of the gravitational waves:~} We now discuss
the propagation of gravitational waves. We introduce tensor-type perturbations
as $g_{ij}=a^{2}(h_{+}\varepsilon_{ij}^{+}+h_{\times}\varepsilon_{ij}^{\times})$,
and $\tilde{g}_{ij}=\tilde{a}^{2}(\tilde{h}_{+}\varepsilon_{ij}^{+}+\tilde{h}_{\times}\varepsilon_{ij}^{\times})$,
with ${\rm tr}(\varepsilon^{+}\varepsilon^{+})=1={\rm tr}(\varepsilon^{\times}\varepsilon^{\times})$,
and ${\rm tr}(\varepsilon^{+}\varepsilon^{\times})=0$. % we can write
%\begin{eqnarray*}
%{\cal L} & = & \frac{\Mpl^{2}a^{2}}{8}\sum_{\lambda={+}}^{\times}\!\left[\dot{h}_{\lambda}^{2}-k^{2}h_{\lambda}^{2}-m^{2}a^{2}\Sigma h_{\lambda}^{2}\right]\nonumber \\
% &  & {}+\frac{\Mpl^{2}a^{2}\kappa\xi^{2}}{8\tilde{c}}\sum_{\lambda={+}}^{\times}\!\left[\dot{\tilde{h}}_{\lambda}^{2}-\tilde{c}^{2}k^{2}\tilde{h}_{\lambda}^{2}-\frac{\tilde{c}\Sigma}{\kappa\xi^{2}}m^{2}a^{2}\tilde{h}_{\lambda}^{2}\right]\nonumber \\
% &  & {}+\frac{\Sigma}{4}\,(a^{4}m^{2}\Mpl^{2})\,\sum_{\lambda={+}}^{\times}h_{\lambda}\tilde{h}_{\lambda}\,.
%\end{eqnarray*}
%where 
%\begin{equation}
%\Sigma\equiv\xi\,[c_{{1}}+2c_{{2}}(1+\tilde{c})\,\xi+6\tilde{c}\,
% c_{{3}}\,\xi^{2}]\,.
%\nonumber
%\end{equation}
The gravitational waves propagate at the speed of light for the physical
sector, whereas at the speed $\tilde{c}\approx1+O(H^{2}/m^{2})$ for
the hidden sector. %ADF: changed on -> for
However, the physical and hidden gravitons, because of the coupling
through the mass term, will oscillate from one to the other. Keeping
only the leading effect of the %ADF: added "the"
deviation of $\tc$ from unity, and neglecting the cosmic expansion
effects, we write the propagation equations as~\cite{Comelli:2012db}
\begin{eqnarray}
 &  & \ddot{h}-\triangle h+m^{2}\Gamma_{c}(h-\tilde{h})=0,\\
 &  & \ddot{\tilde{h}}-\tilde{c}^{2}\triangle\tilde{h}+\frac{m^{2}\Gamma_{c}}{\kappa\xi_{c}^{2}}(\tilde{h}-h)=0\,,\label{GWeqs}
\end{eqnarray}
where we have omitted the $+/\times$ index. %ADF: small change 
For this set of equations, we write down the dispersion relation,
assuming $\tilde{c}-1\ll1$ %ADF: changed as \tilde c > 1
but the magnitude of 
\begin{eqnarray*}
x\equiv\frac{2(2\pi f)^{2}(\tilde{c}-1)}{\mu^{2}}\,,
\end{eqnarray*}
is moderate, where we have defined %ADF: small change
\[
\mu^{2}\equiv\lamg^{-2}=\frac{(1+\kappa\xi_{c}^{2})\,\Gamma_{c}\, m^{2}}{\kappa\xi_{c}^{2}}\,.
\]
Then, for a given gravitational wave frequency $f$, two eigen wave
numbers are given by 
\begin{eqnarray*}
 &  & \!\!\!\!\!\!\!\!\!\! k_{1,2}^{2}=(2\pi f)^{2}-\frac{\mu^{2}}{2}\left(1+x\mp\sqrt{1+2x\frac{1-\kappa\xi_{c}^{2}}{1+\kappa\xi_{c}^{2}}+x^{2}}\right)\,,
\end{eqnarray*}
and the corresponding eigen functions $h_{1}$ and $h_{2}$ are related
to $h$ and $\tilde{h}$ as 
\begin{eqnarray*}
h_{1} & = & \cos\theta_{g}\, h+\sin\theta_{g}\sqrt{\kappa}\xi_{c}\,\tilde{h},\\
h_{2} & = & -\sin\theta_{g}\, h+\cos\theta_{g}\sqrt{\kappa}\xi_{c}\,\tilde{h},
\end{eqnarray*}
with the mixing angle 
\[
\theta_{g}=\frac{1}{2}\cot^{-1}\left(\frac{1+\kappa\xi_{c}^{2}}{2\sqrt{\kappa}\xi_{c}}x+\frac{1-\kappa\xi_{c}^{2}}{2\sqrt{\kappa}\xi_{c}}\right)~.
\]
We find that $\mu$ is the graviton mass of the second mode in the
Minkowski limit ($x\to0$).

When we consider the propagation over a distance $D$, the phase shifts,
due to the modified dispersion relation for their respective modes,
are given by 
\begin{eqnarray*}
\delta\Phi_{1,2}=-\frac{\mu D\sqrt{\tilde{c}-1}}{2\sqrt{2x}}\left(1+x\mp\sqrt{1+x^{2}+2x\frac{1-\kappa\xi^{2}}{1+\kappa\xi^{2}}}\right)~.
\end{eqnarray*}
Notice that this factor is symmetric under the replacement $x\to1/x$.
In the limit $x\to0$, the first mode becomes massless. Although this
mode also has non-trivial dispersion relation, its magnitude of modification
tends to be suppressed. The factor $\mu D\sqrt{\tilde{c}-1}=\sqrt{3(1+\kappa\xi_{c}^{2})\Omega_{0}}HD$
becomes O(1) only after propagating over a cosmological distance unless
$\kappa\xi_{c}^{2}$ is extremely large, where $\Omega_{0}$ is the
energy fraction of the dust matter at the present epoch. On the other
hand, the remaining factor takes the maximum value $2^{-1/2}-(2+2\kappa\xi^{2})^{-1/2}$
at $x=1$, which is also at most $O(1)$. In contrast to the first
mode, the phase shift of the second mode can be significantly large
when $x$ is small or large.
Here we plot $\delta\Phi_{1,2}$ in Fig.~1 for $\kappa\xi_{c}^{2}=0.2,1$
and 100.

%%%%%%%%%%%%%%%%%%%%%%%%%%%%%%%%%%%%%%%%%%%%%%%%%%%%%%%%%%%%%%%%%%%%%%%%%%
%%%%%%%%%%%%%%%%%%%%%%%%%%%%%%%%% Figure 1 %%%%%%%%%%%%%%%%%%%%%%%%%%%%%%%
%%%%%%%%%%%%%%%%%%%%%%%%%%%%%%%%%%%%%%%%%%%%%%%%%%%%%%%%%%%%%%%%%%%%%%%%%%
\begin{figure}
\begin{centering}
\includegraphics[bb=0 0 558 258,width=12cm]{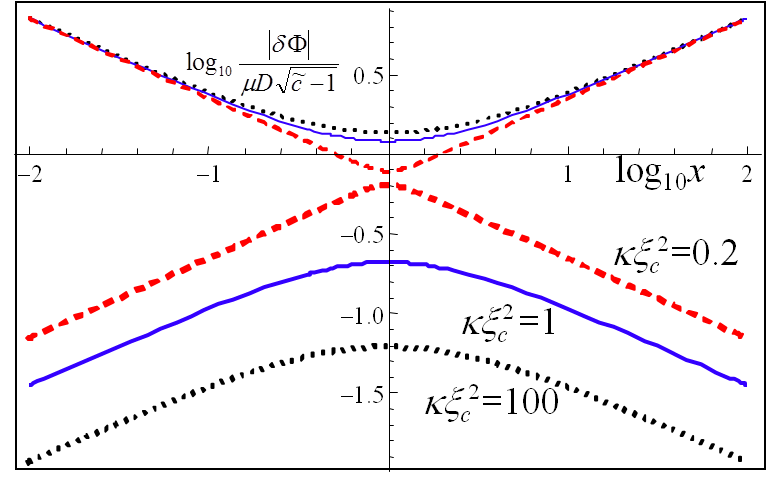} 
\par\end{centering}

\caption{ $|\delta\Phi_{1,2}|$ as a function of $x$ for $\kappa\xi_{c}^{2}=0.2$
(dotted, black), 1 (blue) and 100 (dashed, red). Thick and thin curves
represent $|\delta\Phi_{1}|$ and $|\delta\Phi_{2}|$, respectively. }

%\label{fig1}
\end{figure}

%%%%%%%%%%%%%%%%%%%%%%%%%%%%%%%%%%%%%%%%%%%%%%%%%%%%%%%%%%%%%%%%%%%%%%%%%%
%%%%%%%%%%%%%%%%%%%%%%%%%%%%%%%%%%%%%%%%%%%%%%%%%%%%%%%%%%%%%%%%%%%%%%%%%%

\textbf{Gravitational potential around a star in the Minkowski limit:~}
In the above, we find that, unless $\kappa\xi_{c}^{2}$ is extremely
large, a relatively small value of $\lamg$ together %ADF: added "together"
with the excitation of the second mode is required for an %ADF:added "an"
observable magnitude of the phase shifts due to the non-trivial dispersion
relation. Here we show that in the present bi-gravity models even
with such a small value of $\lamg$ we can easily evade the solar
system constraint from the precision measurement of gravity.

In the low energy limit, it would be natural to assume the hierarchy,
$k^{2}\gg\mu^{2}\gg H^{2}$. Since the limit $H\to0$ is smooth, the
$H$-dependent terms in the action appear as a positive power in $H$.
Since such terms will not give any dominant contribution under the
assumption of the above hierarchy, we set $H=0$ from the beginning
here.

Let us now consider static spherical symmetric perturbations for both
metrics induced by non-relativistic matter energy density $\rho_{{\rm m}}$,
which is coupled only to the physical metric. We can write the respective
perturbed metrics as 
\begin{eqnarray*}
ds^{2} & = & -e^{u-v}dt^{2}+e^{u+v}(dr^{2}+r^{2}d\Omega^{2}),\\
d\tilde{s}^{2} & = & -\xi_{c}^{2}e^{\tilde{u}-\tilde{v}}dt^{2}+\xi_{c}^{2}e^{\tilde{u}+\tilde{v}}(d\tilde{r}^{2}+\tilde{r}^{2}d\Omega^{2}),
\end{eqnarray*}
without loss of generality. Here $\tilde{r}$ is related to $r$ by
$\tilde{r}=e^{{\cal R}}(r)r$, and ${\cal R}(r)$ is another perturbation
variable. We adopted the parametrization such that $u$ vanishes in
the case of general relativity. Now we write down the equations of
motion and eliminate the variables on the hidden metric side, $\tilde{u}$,
$\tilde{v}$ and ${\cal R}$. However, doing this is not so straightforward.
In order to simplify the manipulation, we truncate the perturbation
equations at second order and also neglect higher order terms in $\mu$
appropriately.

When we compute the terms second order in perturbation, we notice
that there are terms enhanced by the factor $1/\mu^{2}$. If we scrutinize
these terms, some of them contain the factor 
\[
C\equiv\left.\frac{d(\log\Gamma)}{d\log\xi}\right\vert _{\xi=\xi_{c}}.
\]
Our assumption here is that the energy scale of the bi-gravity theory
itself is relatively high but the graviton mass $\mu$ is suppressed
by a certain mechanism. %\footnote{A hypothetical 
%braneworld setup may give such an example. Suppose that in 
%the extra-dimensions 
%there are two bulks that are mutually almost disconnected 
%but interact with each other through a narrow throat. 
%In this case, the corresponding low 
%energy effective theory will have two metrics associated with 
%the respective bulks. The energy scale will be determined by the 
%bulk geometry, which will be sufficiently compact, and hence 
%the energy scale $m$ should be high. However, since the link 
%between two bulks is weak, the first excited state of the graviton 
%will have an exponentially suppressed mass. In such a setup, we will 
%have one massless graviton and one massive graviton with an extremely 
%small mass. Such a situation might be realized by choosing the 
%parameters so that $\Gamma\ll 1$ is satisfied but 
%$d\Gamma/d\log\xi|_{\xi=\xi_c}$ is not suppressed at all.}.  
Under this assumption, we pick up only the terms enhanced by the factor
$C/\mu^{2}$ from the second order terms in the equations of motion.
Then, after a little calculation, we obtain 
\begin{eqnarray}
 &  & \!\!\!\!\!\!\!\!\!(\triangle-\mu^{2})u-{\frac{3[(\triangle u)^{2}-(\partial_{i}\partial_{j}u)^{2}]}{8}}\,\frac{\bar{C}}{\mu^{2}}=\frac{\kappa\xi_{c}^{2}\rho_{{\rm m}}}{3\tilde{M}_{G}^{2}}\,,\label{eq1}\\
 &  & \!\!\!\!\!\!\!\!\!\triangle v+3\triangle u-{[(\triangle u)^{2}-(\partial_{i}\partial_{j}u)^{2}]}\frac{\bar{C}}{{\mu}^{2}}=-\frac{\rho_{{\rm m}}}{\tilde{M}_{G}^{2}}\,,\label{eq2}
\end{eqnarray}
where $\partial_{i}$ is the differentiation with respect to the coordinates
$r(\sin\theta\cos\phi,\sin\theta\sin\phi,\cos\theta)$ and $\triangle\equiv\partial_{i}\partial^{i}$,
the standard three dimensional Laplacian operator, and we have defined
$\bar{C}\equiv C(1+\kappa\xi_{c}^{2})/(\kappa\xi_{c}^{2})$. At this
level, the expressions were recast into the form that does not assume
spherical symmetry, where there is no ambiguity.

Although we have truncated the equations at second order for simplicity,
the higher order terms will not be suppressed once the second order
terms become important. %Therefore, we need further study if
%we want to construct the explicit and accurate form of the metric
%perturbation around a star. 
Nevertheless, such higher order terms will not change the following
discussion as to the order of magnitude estimate on the correction
to the Newton's law.

First we focus on Eq.~(\ref{eq1}). Notice that non-vanishing $u$
is the origin of the vDVZ discontinuity~\cite{vDVZ}. This equation
tells us that the Vainshtein radius~\cite{Vainshtein}, within which
the second term dominates the first term on the left hand side in
Eq.~(\ref{eq1}), is given by 
\[
r_{V}=O((Cr_{g}\lamg^{2})^{1/3})\,,
\]
where $r_{g}$ is the gravitational radius of the star. From the above
estimate, we find that the Vainshtein radius can be made arbitrarily
large even with a large graviton mass, if $C$ is sufficiently large.
Thus, the solar system can be easily contained within the Vainshtein
radius, where the second or even higher order terms on the left hand
side of Eq.~(\ref{eq1}) dominate. Then, we have $u\leq O\left([{\kappa\xi_{c}^{2}}/{(1+\kappa\xi_{c}^{2})}]\sqrt{{r_{g}r}/{C\lamg^{2}}}\right)$.
Even if we require $u$ to be smaller than $10^{-9}$ in the solar
system $r\approx10^{13}$cm, $\lamg^{2}$ can be left arbitrarily
small depending on the value of $C$~\cite{Will:2001mx}.

Once $u$ is suppressed on the scale of the solar system, Eq.~(\ref{eq2})
tells us that the equation for $v$ does not largely deviate from
the one in the Newtonian case: $\triangle v=-\tilde{M}_{G}^{-2}{\rho_{{\rm m}}}$,
and the gravitational %ADF: typo
constant is not different from the cosmological one. In Eq.~(\ref{eq2})
the terms second order in $u$ are eliminated with the aid of Eq.~(\ref{eq1})
to make the effective gravitational source for $v$ to be manifest.
Solving the equation, we find that the correction to $v$ is at most
of $O(u)$. Notice that the mass term for $v$ in Eq.~(\ref{eq2})
is absent. Therefore, $v$ does not suffer %ADF
 from the Yukawa-type correction. % which is
%naively expected when the graviton is massive.

Hence, %ADF
 one can conclude that the correction to the Newtonian potential is
at most $O\!\left(\sqrt{\kappa\xi_{c}^{2}r_{g}r/C\lamg^{2}}\right)$.
Namely, we can avoid the constraint from the test in the solar system,
keeping the graviton mass sufficiently large. In the above we assumed
that $C$ is large. On the other hand, in constructing the cosmological
background we have used the %ADF13:added "have" and "the"
linear approximation for the deviation from the
conformal equivalence between the %ADF13:added "the"
two metrics, i.e. $\tilde{c}-1\ll1$.
If we further expand the background equations in terms of $\tilde{c}-1$,
we find terms enhanced by the factor $C$ at second order. However,
as long as $C\lambda_{\mu}^{2}<H^{-2},$ is satisfied, we can verify
that the formul\ae\ for the background metric remain approximately valid.
In the early universe, where $H$ is larger, the non-linear terms
become necessarily important. However, the terms second order in $\xi-\xi_{c}$
do not alter the effective Newton constant for the homogeneous background
cosmology. %in the effective Friedmann equation (\ref{eq:fried1}),
%as long as spatial homogeneity of the universe is assumed.

{The equations for $\tilde{u}$ and $\tilde{v}$ can be obtained
similarly as 
\begin{eqnarray*}
\tilde{u}=-\frac{u}{\kappa\xi_{c}^{2}}\,,\qquad
\tilde{v} = {}v+\frac{3(1+\kappa\xi_c^2)}{\kappa\xi_c^2}u\,.
%-\frac{3[(\triangle
%u)^{2}-(\partial_{i}\partial_{j}u)^{2}]}{2}\frac{\bar C}{\mu^4}\right\}
%\label{eq:vt2-1}
%\tilde{v} & = & -{\frac{3u'\,\left(1+{\xi_{{c}}}^{2}\kappa\right)^{2}\left(u'+2\, ru''\right)C}{{\xi_{{c}}^{4}}{\kappa}^{2}{r}^{2}{\mu}^{4}}}-{\frac{4}{3}}{\frac{\rho}{\Mpl^{2}\mu^{2}}}\nonumber \\
% &  & {}+v-{\frac{(1+{\xi_{{c}}}^{2}\kappa)\, u}{{\xi_{{c}}}^{2}\kappa}}+{\frac{4(1+{\xi_{{c}}}^{2}\kappa)\,\triangle u}{{\xi_{{c}}}^{2}\kappa\mu^{2}}}\,.\label{eq:vt2-1}
\end{eqnarray*}
Once $u$ is suppressed, \textit{i.e.}\ %ADF
if the Vainshtein mechanism is at work, we find $\tilde{v}\approx v$,
%ADF
which implies that metric perturbations on both sides are equally
excited by the matter fields.} %Namely, if we neglect the small difference suppressed as $u$ or $\tilde{u}$,
%a star excites both the physical and hidden metrics equally. In the
%Vainshtein radius the mode that is asymmetric between physical mode
%and hidden mode is suppressed.

\textbf{Graviton oscillations and inverse chirp signal:~} %\textbf{Gravitational wave generation:}~
Here we begin with discussing the generation of gravitational waves.
We found that the metric excitations are almost conformal within the
Vainshtein radius of a star. If we consider the junction between the
near-zone metric perturbation with the far-zone metric described as
gravitational %ADF:typo
waves, both $h$ and $\tilde{h}$ are excited exactly as in the case
of general relativity. This implies that both eigen modes $h_{1}$
and $h_{2}$ are excited unless $x=0$. (Recall that $h_{2}\propto h-\tilde{h}$
when $x=0$.)

%%%%%%%%%%%%%%%%%%%%%%%%%%%%%%%%%%%%%%%%%%%%%%%%%%%%%%%%%%%%%%%%%%%%%%%%%%
%%%%%%%%%%%%%%%%%%%%%%%%%%%%%%%%% Figure 2 %%%%%%%%%%%%%%%%%%%%%%%%%%%%%%%
%%%%%%%%%%%%%%%%%%%%%%%%%%%%%%%%%%%%%%%%%%%%%%%%%%%%%%%%%%%%%%%%%%%%%%%%%%
\begin{figure}
\begin{centering}
\includegraphics[bb=0 0 558 258,width=12cm]{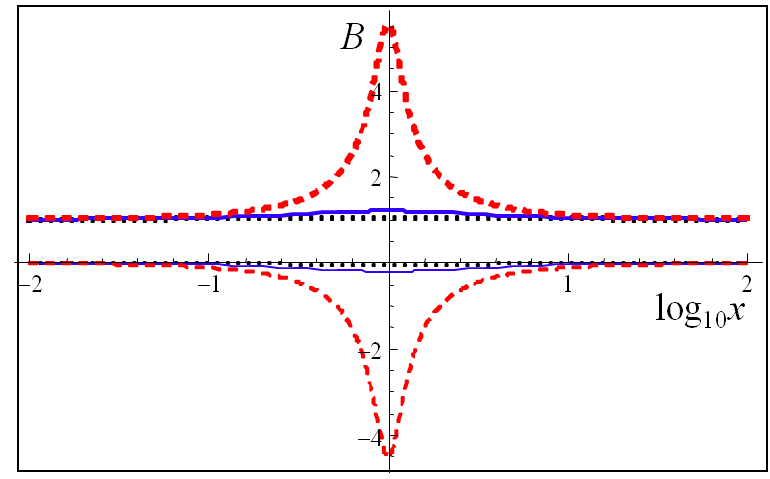} 
\par\end{centering}

\caption{ $B_{1,2}$ as a function of $x$ for $\kappa\xi_{c}^{2}=0.2$ (dotted,
black), 1(blue) and 100(dashed, red). Thick and thin curves represent
$B_{1}$ and $B_{2}$, respectively. }

%\label{fig2}
\end{figure}

%%%%%%%%%%%%%%%%%%%%%%%%%%%%%%%%%%%%%%%%%%%%%%%%%%%%%%%%%%%%%%%%%%%%%%%%%%
%%%%%%%%%%%%%%%%%%%%%%%%%%%%%%%%%%%%%%%%%%%%%%%%%%%%%%%%%%%%%%%%%%%%%%%%%%

One may suspect that the linear approximation to the gravitational
wave perturbation equations (\ref{GWeqs}) is not valid within the
Vainshtein radius. However, the effective energy momentum tensor coming
from the variation of the mass term, which gives corrections to the
case of general relativity, is largely enhanced only for the terms
purely composed of $u$ (or equivalently $\tilde{u}$), which behave
as clouds around localized matter sources. Namely, it just contributes
as the source of gravitational waves but does not change the wave
propagation. The other corrections are suppressed as long as the amplitude
of the deviation of the metric from the case of general relativity
remains small.

Next, we analyze the gravitational waveform from inspirals of NS-NS
binaries at a distance. %In Fig.~1, for various Compton lengths $\lamg$, 
%we show the time evolution of the observed 
%frequency of detector signals for a 1.4$M_\odot+$1.4$M_\odot$  
%NS-NS binary at 300Mpc, when a single massive graviton is assumed.  
%The curve for $\lamg=\infty$ (massless mode) is 
%corresponding to the case of GR. 
%For sufficiently small $\lamg$, we can see the inverse chirp signal 
%due to the slow propagation of massive modes. 
For the current bi-gravity model, our detector signal becomes a linear
combination of two components, whose relative amplitudes are determined
by the mixing angle $\theta_{g}$. For simplicity, we here neglect
the time dependence of $\theta_{g}$ as well as all the cosmological
effects. Using the stationary phase approximation and flux conservation,
the observed signal is given in Fourier space as 
\begin{eqnarray}
h(f)=A(f)e^{i\Phi(f)}\left[B_{1}e^{i\delta\Phi_{1}(f)}+B_{2}e^{i\delta\Phi_{2}(f)}\right]~,\label{vf}
\end{eqnarray}
where the amplitude $A(f)$ (after angular average), $B_{1,2}$ and
the phase function $\Phi(f,g)$ (truncated at 1.5PN order) are given
by 
\begin{eqnarray*}
A(f) & = & \sqrt{\frac{\pi}{30}}\frac{\cM^{2}}{D}u^{-7/6},\\
B_{1} & = & \cos\theta_{g}(\cos\theta_{g}+\sqrt{\kappa}\xi_{c}\sin\theta_{g}),\\
B_{2} & = & \sin\theta_{g}(\sin\theta_{g}-\sqrt{\kappa}\xi_{c}\cos\theta_{g}),\\
\Phi(f) & \equiv & 2\pi ft_{c}-\Phi_{c}-\pi/4+\frac{3}{128}y^{-5/3}\\
 &  & \hspace{-1cm}+\frac{5}{96}\lmk\frac{743}{336}+\frac{11}{4}\eta\rmk\eta^{-2/5}y^{-1}-\frac{3\pi}{8}\eta^{-3/5}y^{-2/3}\,,
\end{eqnarray*}
with $y\equiv\pi\cM f$, the chirp mass $\cM\equiv(m_{1}m_{2})^{3/5}/(m_{1}+m_{2})^{1/5}$
and the reduced mass ratio $\eta={m_{1}m_{2}}/{(m_{1}+m_{2})^{2}}$.
The first and second terms in Eq.~(\ref{vf}) show the contributions
of $h_{1}$ and $h_{2}$, respectively. Here we plot $B_{1,2}$ in
Fig.~2 for $\kappa\xi_{c}^{2}=0.2,1$ and 100.

%%%%%%%%%%%%%%%%%%%%%%%%%%%%%%%%%%%%%%%%%%%%%%%%%%%%%%%%%%%%%%%%%%%%%%%%%%
%%%%%%%%%%%%%%%%%%%%%%%%%%%%%%%%% Figure 3 %%%%%%%%%%%%%%%%%%%%%%%%%%%%%%%
%%%%%%%%%%%%%%%%%%%%%%%%%%%%%%%%%%%%%%%%%%%%%%%%%%%%%%%%%%%%%%%%%%%%%%%%%%
\begin{figure}[t]
\begin{centering}
\includegraphics[bb=0 0 558 268,width=12cm]{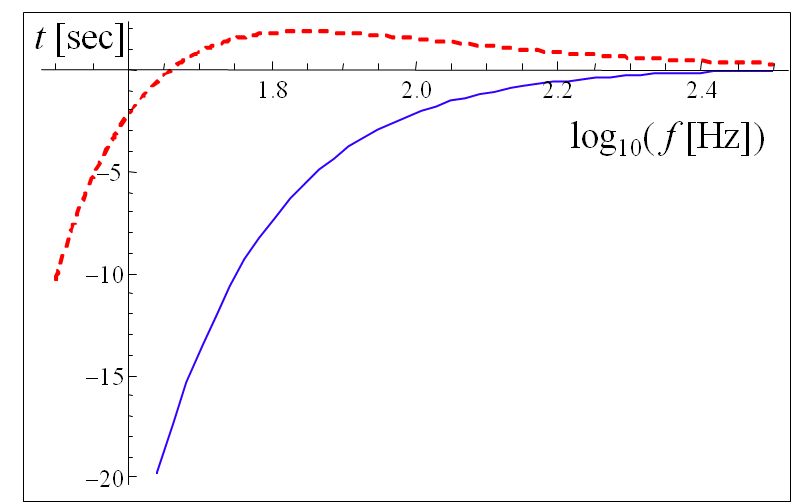} 
\par\end{centering}
\caption{ 
%The shifts of the arrival time as functions of the frequency
%$f$ for respective modes %ADF:typos
%compared with the case of general relativity 
The arrival time as functions of the frequency $f$ for
respective modes %ADF:typos
for $1.4M_{\odot}+1.4M_{\odot}$
binary inspiral with $\kappa\xi_{c}^{2}=100$, $D=300$Mpc $H=67.3$km~s$^{-1}$Mpc$^{-1}$,
$\Omega_{0}=0.315$ and $\lambda_{\mu}=0.001{\rm pc}$. 
The blue solid curve is for the first mode, while the dashed 
red one for the second mode. 
}
%\label{fig3}
\end{figure}

{
For $x\ll1$, the excitation of the second mode $h_{2}$ is suppressed.
Furthermore, $\delta\Phi_{1}$ is suppressed in this regime. Therefore,
the propagation of gravitational waves is similar to the case of general
relativity. For $x\gg 1$, both $h_{1}$ and $h_{2}$ are equally excited. 
However, since the gravitational wave detector can detect the 
perturbation of the physical metric only, we can observe only $h_{1}$. 
Therefore, the frequencies at which both modes can be observed are limited 
to $x\approx 1$. This is the meaning of Fig.~2. 

When both modes are observable, 
graviton oscillations due to the interference between two modes 
can be detected beyond the distance scale where
$\delta\Phi_{2}-\delta\Phi_{1}$ becomes $O(1)$. 
The difference of the phases 
$\delta\phi_{1}-\delta\phi_2$ 
is minimum at $x= 1$ as shown in Fig.~1, 
which is evaluated as $\delta\phi_{1}-\delta\phi_2|_{x=1}
=\sqrt{6\Omega_0}HD$.  
Therefore, one may think that the effect is really small 
as long as $D\lesssim H^{-1}$. 
However, the average density of the universe is much lower 
than the average density in galaxies, where binaries are 
embedded. Therefore gravitational waves experience 
much lower value of $x$, typically $x\approx 10^{-8}$,  
during the propagation. Roughly speaking,
$\delta\Phi_{2}-\delta\Phi_{1}\approx\sqrt{3(1+\kappa\xi_c^2)\Omega_0/2x}H
D$ for $x\ll 1$. Hence, the effect can be largely enhanced. 

Once $d(\delta\Phi_{2}-\delta\Phi_{1})/df$ becomes sufficiently large,
the arrival times of two modes are different. Then, we may observe two
chirp signals. Using the stationary phase
approximation, the relation between the arrival time of the wave and
the frequency is determined by $t={d(\Phi+\delta\Phi_{i})}/{d(2\pi f)}$.
As an illustrative purpose, in Fig.~3 we show the shifts of the
arrival time compared with the case of general relativity for 
$\kappa\xi_c^2=100$,
$D=300$Mpc $H=67.3$km~s$^{-1}$Mpc$^{-1}$, $\Omega_{0}=0.315$ and
$\mu=(0.001{\rm pc})^{-1}$, for which $x\approx 4\times 10^{-8}$ at 
100Hz. 
% If we further reduce the graviton mass,
%e.g. $\lambda_{\mu}=0.001$pc, as is shown in fig3.(b) 
One can see that the relation
between the arrival time and the frequency is reversed for the second
mode, i.e. inverse chirp signal may occur.
}

Since a large graviton mass can be consistent with the solar system
test in our current model%
\footnote{In fact, another possibility of suppressing $u$ in Eq.~(\ref{eq1})
is to consider a really large value such as $\mu^{-1}<0.1$mm, In
this case the massive graviton will be almost irrelevant to describe
the present-day gravity, but it will alter the cosmology in the early
epoch. This possibility might be explored elsewhere.%
}, there is a possibility that we may detect the graviton oscillations
or even the inverse chirp signal by the next generation gravitational
wave detectors. In the present model, measurable effects are expected
only when $\kappa\xi_{c}^{2}$ is large. This requirement may cause some conflict
with observations but at first analysis there seems to
be no severe constraint. We think this model gives the first existence
proof of models in which a measurable deviation from general relativity
in the gravitational wave propagation can be expected.
\begin{acknowledgments}
We would like to thank Naoki Seto, Michael Volkov and Sergey 
Sibiryakov for their valuable comments. This work
is supported by Monbukagakusho Grant-in-Aid for Scientific Research
Nos.~24103006, 21244033, 21111006, 23540305 and the Grant-in-Aid
for the Global COE Program ``The Next Generation of Physics, Spun
from Universality and Emergence'' from the Ministry of Education,
Culture, Sports, Science and Technology (MEXT) of Japan.

%\cite{Hassan:2011zd}

%\cite{Will:1997bb}

\end{acknowledgments}

\end{document}